\documentclass[fleqn,usenatbib]{mnras}

\usepackage{newtxtext,newtxmath}

\usepackage[T1]{fontenc}

\DeclareRobustCommand{\VAN}[3]{#2}
\let\VANthebibliography\thebibliography
\def\thebibliography{\DeclareRobustCommand{\VAN}[3]{##3}\VANthebibliography}

\usepackage{graphicx}	
\usepackage{amsmath}	

\title[Transient and Variable ULXs in NGC 4552]{Transient and Variable Ultra-luminous X-ray Sources in NGC 4552}

\author[Jithesh et al. 2026]{
V. Jithesh,$^{1}$\thanks{E-mail: jithesh.v@christuniversity.in}
A. S. Sreya,$^{1}$
Riya Elza Mathew,$^{1}$
E. Sreeraj,$^{1}$
and R. Arun$^{1, 2, 3}$
\\
$^{1}$Centre of Excellence in Astronomy and Astrophysics (CEAA), Department of Physics \& Electronics, Christ University, Hosur Main Road, Bangalore 560029, India\\
$^{2}$Instituto de Estudios Astrofísicos, Facultad de Ingeniería y Ciencias, Universidad Diego Portales, Av. Ejército Libertador 441, Santiago, Chile\\
$^{3}$Millennium Nucleus on Young Exoplanets and their Moons (YEMS), Chile\\
}

\date{Accepted XXX. Received YYY; in original form ZZZ}

\pubyear{\the\year{}}

\begin{document}
\label{firstpage}
\pagerange{\pageref{firstpage}--\pageref{lastpage}}
\maketitle

\begin{abstract}
We searched for transient and variable X-ray sources in the elliptical galaxy NGC 4552 using the \textit{Chandra} and  \textit{XMM-Newton} observations from 2001-2012. We detected 14 transient and variable X-ray sources within the 4$R_{e}$ region of the galaxy, which exhibited peak state and undetected flux behaviour  in the \textit{Chandra} observations. Among them, two sources (T1 and T14) exceed the X-ray luminosity $10^{39}\rm~erg~s^{-1}$ in the detected observations and have the peak state and undetected luminosity ratio of $\geq 14.3$ and $\geq 8.7$, respectively, which we classified as transient and variable ultra-luminous X-ray sources (ULXs). For the third relatively bright transient source, the flux varies by a factor of 3 across the detected observations, and the hardness ratio analysis suggests a normal X-ray binary system. The remaining sources have luminosities in the $10^{37}-10^{38} \rm~erg~s^{-1}$ range, and the spectral analysis indicates that the sources may belong to neutron star or black hole X-ray binary systems, either in the hard or thermal-dominated state. The identified transient and variable ULXs exhibited high variability and soft spectral characteristics, which are consistent with ULXs accreting at super-Eddington rates. The near-infrared (NIR) counterpart search using {\it JWST} observations identified potential NIR counterparts for three mildly variable sources, and their magnitudes in different NIR bands suggest a red supergiant companion for each source. Deep X-ray monitoring observations can shed further light on the new transient and variable ULXs.
\end{abstract}




\section{Introduction}

X-ray transients, also known as X-ray novae, are astronomical phenomena characterised by sudden, transient surges of X-ray radiation from celestial objects that typically do not exhibit persistent X-ray emission. X-ray transients have been a subject of extensive study in X-ray astronomy since their discovery, with the majority found in binary systems with a primary component of either a black hole (BH) or a neutron star (NS). The transient population has been well studied in the Milky Way Galaxy, and one-third of known X-ray binaries (XRBs) exhibit transient behaviour \citep{1996ARA&A..34..607T}. Apart from the Milky Way, transient sources have been searched in the Small and Large Magellanic Clouds (SMC and LMC), and many of them are identified as super-soft sources (SSSs) and pulsars \citep{1996A&A...312..919K, 2002A&A...385..464L, 2010A&A...517L...5O, 2013PASJ...65L...2W}. Due to the proximity and moderate Galactic foreground absorption, a large number of studies have been conducted in M31 to search and understand the transient population \citep[][and references therein]{2006ApJ...643..356W, 2006ApJ...645..277T, 2007A&A...468...49V, 2008A&A...480..599S}. Monitoring programs identified several recurring transients, and catalogue studies showed a minimum rate of nine transient sources per year \citep{2008A&A...489..707V}, suggesting that M31 is a transient factory. A sizeable population of eight transients has been identified from M33 galaxy using {\it Chandra} and {\it XMM-Newton} observations \citep{2008ApJ...680.1120W, 2013MNRAS.435.3326T}. In M33, the bright transient sources are consistent with high-mass X-ray binaries (HMXRBs), unlike transients in M31, which are dominated by low-mass X-ray binaries (LMXRBs). In NGC 55, 15 transient sources have been identified \citep{Jithesh_2016}, and the majority are LMXRBs and HMXRBs, the remaining sources are consistent with SSSs. A small population of seven transients in NGC 4449 were identified using the {\it Chandra} observations; two sources were identified as SSSs, and the rest of them are X-ray binaries \citep{Jithesh_2017}. These studies revealed that the transient behaviour exhibited by different categories of sources, including LMXRBs, HMXRBs, SSSs, and Ultra-luminous X-ray sources (ULXs).

ULXs are defined as point-like X-ray sources in the non-nuclear region of a galaxy with an observed luminosity above $10^{39}\rm~erg~s^{-1}$ \citep{kaaret2017ultraluminous, 2021AstBu..76....6F, 2023NewAR..9601672K}. Earlier studies suggested an intermediate black hole, with a mass range of $10^{2} - 10^{5}$\(M_\odot\), as an accretor for ULXs with a sub-Eddington accretion rate \citep{1999ApJ...519...89C, 2004IJMPD..13....1M}, but later high-quality XMM-Newton observations of several ULXs revealed a ubiquitous feature: a soft excess and broad curvature at high energies \citep{2006MNRAS.368..397S, 2009MNRAS.397.1836G}. These features indicate that ULXs are in ``ultraluminous state'', which is powered by super-Eddington accretion onto the stellar mass black hole \citep{2009MNRAS.397.1836G, 2015MNRAS.447.3243M, 2015ApJ...814...73S, 2021MNRAS.504..974G}. In the last decade, a new population of ULXs emerged, which is powered by neutron stars and is known as pulsating ULXs \citep[PULXs;][]{2014Natur.514..202B, 2017Sci...355..817I, 2018MNRAS.476L..45C, 2020ApJ...895...60R}. The majority of ULXs are persistent sources, and a sub-population of ULXs has been identified as transients, where the source exhibited on-off behaviour \citep{2012MNRAS.420.2969M, 2012ApJ...750..152S, 2018MNRAS.477L..90P, 2019MNRAS.483.3566V, 2020ApJ...891..153E, 2020ApJ...890..166P, 2020ApJ...895..127B, 2021MNRAS.501.1002W, 2022MNRAS.511.4528U, 2022MNRAS.515.4669R, 2023ApJ...951...51B, Allak2023, 2024MNRAS.528..418L, 2026A&A...707A.236A}. Since these sources are transitional objects between normal XRBs and bright ULXs, it is important to search for them in external galaxies.   

 NGC 4552 (also known as M89) is an elliptical galaxy in the Virgo cluster located at a distance of 16.25 Mpc \citep{Tonry}. The central source is bright in radio wavelength and has a relatively flat spectrum \citep{1984ApJ...287...41W, 2000ApJS..129...93F}. In addition, the central source showed high variability in the near-UV band \citep{1995Natur.378...39R, 1999fgb..conf..191C}, indicating that the galaxy hosts a mini active galactic nuclei (AGN) at its centre. X-ray point source population studies with Chandra observations \citep{Xu_2005} detected 47 X-ray point sources within four effective radii, and most of them are likely LMXRBs. The brightest X-ray source position is consistent with the optical, UV, and radio centres of the galaxy, which is a low-luminosity AGN that showed variability on timescales of $> 1$ hours. The off-nuclear X-ray sources have a luminosity range extending from 7 $\times 10^{37}$ to 1.5 $\times 10^{39}\rm~erg~s^{-1}$. Among them, three sources have luminosities exceeding $10^{39} ~ \rm erg~s^{-1}$. 

In this paper, we revisit the archival {\it Chandra} and {\it XMM-Newton} observations of NGC 4552 to search for transient and variable X-ray sources and ULXs. We report the spectral and temporal properties of these X-ray sources identified in NGC 4552 and searched for potential near-infrared (NIR) counterparts using the {\it James Webb Space Telescope} ({\it JWST}). The paper is structured as follows: In \ref{sect:Obs}, we describe the observations from the \textit{Chandra}, \textit{XMM-Newton}, and {\it JWST} observatories and the data reduction. We discuss the identification of transient and variable sources and their properties in \ref{sec:analysis}. In \ref{sec:discussion}, we discuss the results of the study and summarise our results.

\section{Observations and Data Reduction}
\label{sect:Obs}

We used the X-ray data of NGC 4552 from \textit{Chandra} \citep{2000SPIE.4012....2W} and \textit{XMM-Newton} \citep{Jan01} observatories during the period 2001 April - 2012 August. We primarily used \textit{Chandra} data for the detection of transient sources, while \textit{XMM-Newton} observation was used for follow-up studies. The details of the observations used are given in Table \ref{Table 1}.

\begin{table}
\begin{center}
\caption[Observation Log]{Observation log of \textit{Chandra }and \textit{XMM-Newton}}
\label{Table 1}
\begin{tabular}{clclclclcl}
     \hline
      Mission & Data & ObsID & Date & Exposure$^a$ \\
    \hline
     \textit{Chandra }& C1 & 2072 & 22-04-2001 & 55.14 \\
      & C2 & 13985 & 22-04-2012 & 50.07 \\
      & C3 & 14359 & 23-04-2012 & 48.77 \\
       & C4 & 14358 & 10-08-2012 & 50.06 \\
       \textit{XMM-Newton} & XMM1 & 0141570101 & 10-07-2003 & 44.8 \\
     \hline
\end{tabular}
\end{center}
     \small
     Note - $^a$Exposure time is in units of kiloseconds.
\end{table}

\subsection{\textit{Chandra}}
We found four sets of archival \textit{Chandra} observations of NGC 4552 spanning 2001--2012. This study used data from the \textit{Chandra} Advanced CCD Imaging Spectrometer-Spectroscopy Array (ACIS-S). We analysed the data using \textit{Chandra} Interactive Analysis of Observation (CIAO; version 4.14) software with \textit{Chandra} Calibration Database (CALDB; version 4.10.4). We reprocessed the data using the ''chandra\_repro'' tool in CIAO. Following the extraction of the exposure map using CIAO, we used the Mexican-hat wavelet source detection tool, WAVDETECT \citep{Freeman2001AWA} in the 0.3--10 keV energy band. In the WAVDETECT tool, we used the wavelet scales of 1.4, 2, 4, 8, and 16 pixels and a detection threshold of $5 \times 10^{-6}$. With this criterion, we detected 93-126 X-ray point sources in the observations we analysed.

\subsection {\textit{XMM-Newton}}

NGC 4552 was observed once with the \textit{XMM-Newton} telescope on 2003 July 10 for an exposure time of 44.8 ks. The observation details are given in Table \ref{Table 1}. We utilised data from the European Photon Imaging Camera (EPIC) PN \citep{refId0} and MOS \citep{2001A&A...365L..27T} detectors for this analysis. The EPIC-PN and EPIC-MOS data were reduced using EPCHAIN and EMCHAIN of \textit{XMM-Newton} Science Analysis Software (SAS; version 20.0). We extracted the particle-flaring background light curve in the 10-12 keV energy range and applied count-rate criteria (0.5 c/s for PN and 0.4 c/s for MOS) to create the good-time interval (GTI) file. Using the GTI file, we extracted the cleaned and filtered event file for further analysis.  We created a detection map using a threshold of 0.25 and ran the SAS task EBOXDETECT in local mode for the initial source detection (with a detection threshold likemin = 8). The obtained source position was used as input to the task ESPLINEMAP, which constructs background maps from source-free regions of the image. The EBOXDETECT task was run again in map mode, using the available background map to improve detection. We ran the EMLDETECT task to perform maximum-likelihood fitting of the sources and obtained the final source list for the observation. 

\subsection {\textit{JWST}}

{\it JWST} \citep{JWST2006} observed NGC 4552 in four filters, F090W (exposure time of 5840.8\,s), F150W (1245.5\,s), F277W (5840.8\,s) and F356W (1245.5\,s), of the NIRcam Imaging module \citep{Rieke2005}, on 2024 June 26 (PI: R. Brent Tully; Program ID~\#3055). We downloaded calibrated mosaic images from the MAST archive \footnote{\url{https://mast.stsci.edu/portal/Mashup/Clients/Mast/Portal.html}} and used them for subsequent analysis.

\section{ANALYSIS AND RESULTS}\label{sec:analysis}
\subsection{Identification of Transient and Variable Sources}
\label{subsec:transient}

\begin{figure}
\includegraphics[scale=0.48]{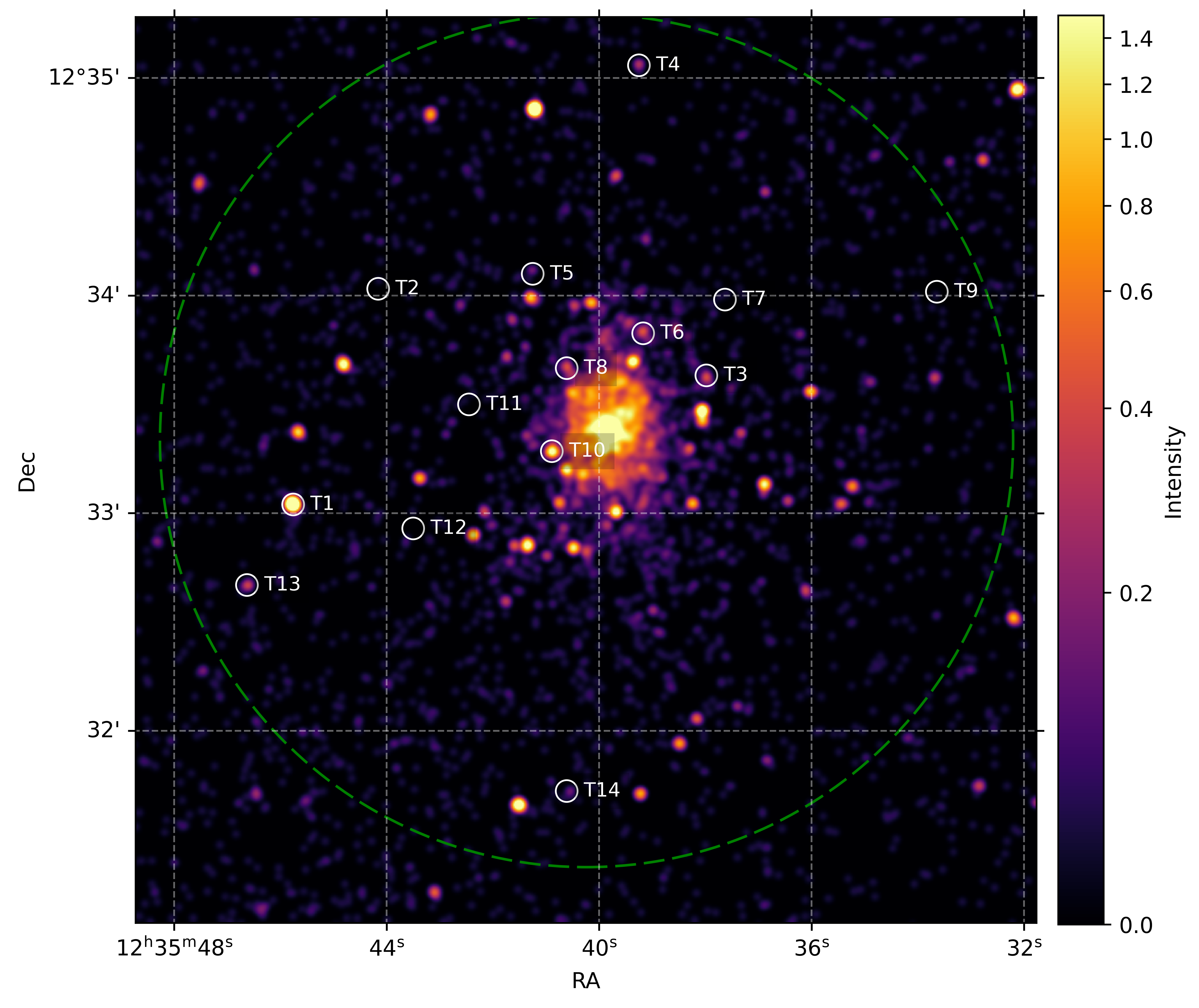}
\caption{Smoothened {\it Chandra} ACIS image of NGC 4552 from C1 observation. The dashed green circle represents the $4R_e$ region of the galaxy. The detected transient sources are represented as solid white circles with a radius of 3 arcseconds.}
\label{fig:Chandra_image_NGC4552}
\end{figure}

Transient and variable X-ray sources were visually identified based on their peak and undetected-intensity states. We used mainly long-separated and long-exposure \textit{Chandra} observations (C1 and C4) to search for transient and variable sources \citep{Jithesh_2016}. The remaining observations were used for the follow-up studies. 126 and 100 X-ray point sources are detected in the C1 and C4 observations, respectively. Of these sources, 64 and 57 are within four effective radii (4$R_{e}$) in C1 and C4 observations, respectively. On visual inspection, we identified 14 X-ray point sources within the 4$R_{e}$ circle (see Figure \ref{fig:Chandra_image_NGC4552}) that showed peak state and undetected flux behaviour. To avoid false detection of these sources, we have put stringent criteria that the source should be detected in at least one of the observations at a $> 4 \sigma$ confidence level with an unabsorbed 0.3--10 keV luminosity of $> 1 \times 10^{36}\rm erg~s^{-1}$. All 14 sources satisfy this criterion. Furthermore, we calculated the ratio of the peak-state to undetected luminosities. For sources located in distant galaxies, such as NGC 4552, given the current sensitivity limits of our X‑ray telescopes, it is quite hard to detect the quiescent state. Perhaps for this reason, we used the following criteria and terminology to classify the X-ray sources. We classified the sources with a luminosity ratio $>10$ as transient sources (TSs). If the sources have ratios between 5 and 10, we classify them as variable sources (VSs); ratios $< 5$ are classified as mildly variable sources (MVSs). This scheme identifies two TSs and one VS, with the remaining being MVSs. The details are given in Table \ref{Table 2}. Among the 14 sources, three (T1, T2 and T14) sources were detected in the 2003 \textit{XMM-Newton} observation. 


\begin{table*}
    \centering
    \setlength{\tabcolsep}{8.0pt}
    \caption{Transient and variable sources detected in NGC 4552}
     \label{Table 2}
    \begin{tabular}{cccccccccc}
    \hline
Source &  Obs ID  &R.A.& Decl. & Positional & Net  & $L_{{x}_{upper}}$ & Luminosity & Class\\
   &   &(hh:mm:ss)&(\textsuperscript{o}: $\prime$ : $\prime\prime$)  & Uncertainty ('') & Counts & ($\rm erg~s^{-1}$) & Ratio & \\
        \hline
       T1 &  C1 &  12:35:45.76 & +12:33:02.4 & 0.04 & 363 & $1.76 \times 10^{38}$
& $\geq$ 14.28 & TS\\
          & XMM1 &   &  & & 83 & \\
    
       T2 &   C2 & 12:35:44.17 & +12:34:01.7  & 0.04 & 138 & $5.63 \times 10^{37}$
 & $\geq$ 15.48& TS \\
          &   XMM1 &  & &  & 110 &  \\
            & C3  &  & & &  140  & \\
            &  C4 & & &  &47 & \\
       
        T3  &    C1  &12:35:37.98 & +12:33:38.0 & 0.13 & 99& $3.34 \times 10^{38}$
& $\geq$ 0.97 & MVS\\
            &   C3 & & & & 51 & \\
            &   C4 &  & & & 34 & \\
        T4 &    C1& 12:35:39.25 & +12:35:03.5 & 0.16 & 29 & $3.09 \times 10^{38}$
& $\geq$ 0.36 & MVS \\
          &C4 & & & & 31 & \\
        T5 &   C4 & 12:35:41.25 & +12:34:06.0 & 0.28 & 32 & $1.79 \times 10^{38}$
& $\geq$ 0.47 & MVS \\
        T6 &    C1& 12:35:39.17 & +12:33:49.6  & 0.11 & 155 & $6.68 \times 10^{38}$ & $\geq$ 0.73 & MVS\\
             & C2& & & & 110 & \\
        T7 &  C2& 12:35:37.63 & +12:33:58.9 & 0.35 & 35 & $2.52 \times 10^{38}$
 & $\geq$ 0.55 & MVS\\
            &  C3 & &  & & 36\\
        T8 &   C1 & 12:35:40.61 & +12:33:40.0 & 0.12 & 165 & $4.83 \times 10^{38}$
& $\geq$ 0.95 &MVS\\
          &   C2 & & & & 103 & \\
        T9 & C4 & 12:35:33.64 & +12:34:01.6 & 0.45 & 18 & $1.22 \times 10^{38}$
& $\geq$ 0.45 & MVS \\        
        T10 &  C1 & 12:35:40.89 & +12:33:17.1  & 0.06 & 273 & $4.31 \times 10^{38}$
& $\geq$ 2.12& MVS\\
        T11 &  C2 & 12:35:42.45 & +12:33:30.0  & 0.26 & 26 & $1.04 \times 10^{38}$
& $\geq$ 0.35 & MVS \\
        T12 & C2 & 12:35:43.50 & +12:32:55.8 & 0.34 & 72 &  $1.55 \times 10^{38}$
& $\geq$ 0.43 & MVS \\
        T13 &  C1 & 12:35:46.63 & +12:32:40.2  & 0.15 & 34 & $1.41 \times 10^{38}$
 & $\geq$ 0.98 & MVS\\
            &  C2 & & & & 32 & \\
            &  C4 & & & & 26 & \\
        T14 &  C1 & 12:35:40.61 & +12:31:43.4  & 0.19 & 33 & $1.55 \times 10^{38}$
 & $\geq$ 8.69 & VS \\  
            &  XMM1 &  & &  & 237 & \\  
       
    \hline
    \end{tabular}
\begin{flushleft}
    \small 
    Note- 1) The name of the sources referred to in this work, 2) observation ID in which they are detected, 3) right ascension (R.A.), 4) declination (Decl.) of detected sources (J2000), 5) positional uncertainty in arcseconds, 6) net counts in the 0.3-10 keV (in the case of {\it XMM-Newton}, the counts from PN data), 7) strict upper limit on the luminosity in $\rm erg~s^{-1}$, 8) ratio of peak state luminosity to non-detection luminosity, 9) source classification (TS = Transient Source, MVS = Mildly Variable Source, VS = Variable Source).
\end{flushleft}
\end{table*}

\subsection{Estimation of Hardness Ratio}
\label{sec_HR}

Due to the limited net counts of transient and variable sources, the hardness ratio (HR) can be regarded as a primary tool for examining their spectral properties. For \textit{Chandra} data, HRs are defined as HR1 = $\frac{(M - S)}{(M + S)}$ and HR2 = $\frac{(H - M)}{(H + M)}$, where S, M, and H are the count rates in soft (0.3 - 1 keV), medium (1 - 2 keV) and hard (2 - 8 keV) bands, respectively. In the case of \textit{XMM-Newton} data, we calculated HRs using count rates in the 0.3–1 keV (soft), 1–2 keV (medium) and 2–6 keV (hard) bands \citep{Jithesh_2016}. We used the X-ray colour classification scheme given in \citet{2005ApJS..159..214K} and \citet{2005MNRAS.357..401J} for \textit{Chandra} and \textit{XMM-Newton} data, respectively.

The calculated hardness ratios and their classification are listed in Table \ref{Table 3}. The X-ray colour classification scheme divides the X-ray sources into different categories: Supernova remnant (SNR) HR2 $< -0.2$, HR1 $< -0.3$,
X-ray binary (XRB)$ -0.8 < $ HR2 $< 0.8, -0.3 <$ HR1 $< 0.6$,
Background source (BKG) HR2 $> -0.2$, HR1 $< -0.3$,
Absorbed source (ABS) HR1 $> 0.6$, Indeterminate soft source (ISS) HR2 $< -0.8, -0.3 < $ HR1 $< 0.6$, Indeterminate hard source (IHS) HR2 $> 0.8, -0.3 < $ HR1 $ <$ 0.6, Super soft source (SSS) M = 0, H = 0
 (refer Table 2 of \cite{2005ApJS..159..214K} for more details). The X-ray colour classification scheme used in \cite{2005ApJS..159..214K} is based on assumed column density values. Since NGC 4552 exhibits a steep density profile, sources located in the inner regions of the galaxy may experience higher local absorption. Increased column density can modify the observed hardness ratios, potentially affecting the classification of sources, particularly between supersoft sources, supernova remnants, and X-ray binaries. Therefore, the colour-based classifications presented here should be considered indicative rather than definitive for highly absorbed sources.

 \begin{table}
    \centering
    \caption{Hardness ratios of transient and variable sources} 
    \label{Table 3}
    \begin{tabular}{clclclclcl}
    \hline
    Source  & ObsID  & HR1 &  HR2 &  Class  \\
        \hline
    T1 & C1 &  +0.18 $\pm$ 0.05 & -0.52 $\pm$ 0.06& XRB \\
       & XMM1  &  -0.40 $\pm$ 0.28 & -0.29 $\pm$ 0.48 & SNR \\
    T2 & C2 & +0.36 $\pm$ 0.09 & -0.57 $\pm$ 0.11&XRB \\
         &  XMM1  &  -0.34 $\pm$ 0.17 & +0.10 $\pm$ 0.18 & XRB \\
            &C3  & +0.23 $\pm$ 0.10 & -0.25 $\pm$ 0.12 & XRB\\
            & C4 & +0.38 $\pm$ 0.19 & -0.56 $\pm$ 0.26 &XRB\\
    T3  & C1 & -0.34 $\pm$ 0.11 & -0.47 $\pm$ 0.20&XRB\\
            &  C3 & +0.04 $\pm$ 0.23  & -0.13 $\pm$ 0.25 &XRB\\
             & C4 & -0.33 $\pm$ 0.28& -0.25 $\pm$ 0.40 & XRB\\
    T4 &   C1 & +0.20 $\pm$ 0.47 & -0.09 $\pm$ 0.37& XRB\\
            & C4 & +0.70 $\pm$ 0.65& -0.07 $\pm$ 0.31 &ABS\\
    T5 & C4 & +0.59 $\pm$ 0.57 & -0.99 $\pm$ 0.93 &SOFT\\
    T6 & C1 & -0.46 $\pm$ 0.08& -0.94 $\pm$ 0.26 &SNR\\
            &C2& -0.14 $\pm$ 0.11& -0.46 $\pm$ 0.17  & XRB\\
    T7 & C2 & -0.07 $\pm$ 0.38 & -0.71 $\pm$ 0.96 & XRB\\
            & C3 & -0.28 $\pm$ 0.35 & -0.11 $\pm$ 0.50 &XRB\\
    T8 & C1 & -0.56 $\pm$ 0.07 &  -0.54 $\pm$ 0.18 &SNR\\
          & C2 & -0.63 $\pm$ 0.12 & -0.04 $\pm$ 0.30 & SNR\\
    T9 & C4 & -0.14 $\pm$ 0.49 & -0.20 $\pm$ 0.72 &XRB\\
    T10 & C1 & -0.36 $\pm$ 0.06 & -0.51 $\pm$ 0.09 &XRB\\
    T11 & C2 & -0.25 $\pm$ 0.51 & +0.14 $\pm$ 0.58 & XRB \\
    T12 & C2 & -0.66 $\pm$ 0.64 & +0.50 $\pm$ 0.95&XRB\\
    T13 & C1 & +0.46 $\pm$ 0.30 & -0.37 $\pm$ 0.29 &XRB\\
            & C2 &  +0.49 $\pm$ 0.57 & -0.33 $\pm$ 0.55 &XRB\\
            & C4 & -0.19 $\pm$ 0.55  & +0.42 $\pm$ 0.54 &XRB\\
    T14 & C1 & +0.55 $\pm$ 0.44 & -0.07 $\pm$ 0.39 & XRB\\    
         &  XMM1  &  -0.66 $\pm$0.12 & +0.20 $\pm$0.18 & BKG \\
    \hline 
    \end{tabular}
    \begin{flushleft}
    \small 
    Note- 1) Name of the source, 2) the observation ID in which they are detected, 3) the Hardness Ratio derived from the count rates (See Section \ref{sec_HR}), and 4) the nature of the sources according to the classification scheme of \cite{2005ApJS..159..214K} and \cite{2005MNRAS.357..401J}.
    \end{flushleft}
\end{table}

\subsection{Spectral Analysis of Transient and Variable Sources}

X-ray spectra of transient and variable sources within the 4$R_{e}$ region were extracted in the 0.3–10 keV band. For the {\it Chandra} observations, we extracted the source events from a circular region of radius 5 arcseconds. The background spectra were extracted from the nearby source-free region with the same radius as the source. We used the CIAO tool \textit{specextract} to obtain the source and background spectra together with the Ancillary Response File (ARF) and Redistribution Matrix File (RMF). We binned the spectra with at least 20 counts per bin using the GRPPHA tool in XSPEC version 12.12.1 \citep{1996ASPC..101...17A}. For the spectrum grouped with a minimum of 20 counts per bin, we used the $\chi^2$ statistic for spectral fitting. Some sources have a few counts below 100 counts. For those cases, we used the Cash statistics \citep{1979ApJ...228..939C} for spectral fitting. The quality of the fit in the C-statistics was obtained by performing 1000 Monte Carlo simulations using the GOODNESS task \citep{1996ASPC..101...17A}. The fit is accepted if the "goodness" is $\leq$ 50 $\%$.

The follow-up study of transient and variable sources is performed using the sole {\it XMM-Newton} observation. Only three transient sources (T1, T2, and T14) are detected in this observation. We extracted the PN and MOS spectra of these sources with a source radius of 13-15 arcsecs. The selection of these source region radii is to avoid the CCD gap and edge. The background regions are selected from the same CCD, with a radius equal to that of the source region. We extracted the source and background spectra along with the ARF and RMF. We grouped the spectra of these sources with a minimum of 20 counts per bin using {\sc grppha} tool and fitted the PN and MOS spectra jointly.

In the detected cases, we modelled the transient source spectra using an absorbed disk blackbody model (TBABS*DISKBB) and an absorbed power law model (TBABS*PL). The absorption column density in the TBABS model is fixed at N$_H$ = 2.67$\times$10\textsuperscript{20} cm\textsuperscript{-2} \citep{2016A&A...594A.116H}, and the Solar Abundance Vector set to angr \citep{1989GeCoA..53..197A}. The unabsorbed flux is estimated using the convolution model CFLUX available in XSPEC, and the X-ray luminosities were calculated assuming a distance of 16.25 Mpc \citep{Tonry}. All errors quoted were computed at a confidence level of 90$\%$. When using the $\chi^2$ statistic, the lowest $\chi^2$ model is chosen as the preferred model.

\begin{table*}
    \centering
    \setlength{\tabcolsep}{4.0pt}
    \caption{Best fit parameters of the transient and variable sources in NGC 4552}
    \begin{tabular}{cccccccccccc}
    \hline
        &&&&$PL^{a}$&&&&$DISKBB^{a}$& \\
        Source$^{b}$ & ObsID$^{c}$ & $\Gamma^{d}$ & $\rm Norm_{PL}^{e}$ & $log L_{X}^{f}$ & $\chi^2$/d.o.f$^{g}$ & G$^{h}$ & kT$_{in}^{i}$& $\rm Norm_{Diskbb}^{j}$ & log $L_{X}^{f}$& $\chi^2$/d.o.f$^{g}$& G$^{h}$\\
     \hline
         T1  & C1 & $2.67^{+0.49}_{-0.41}$ &$(1.66^{+0.73}_{-0.48})\times 10^{-5}$ & $39.40^{+0.20}_{-0.14}$& 12.36/13 & 0.49 &  $0.82^{+0.13}_{-0.11}$ & $(3.31^{+2.40}_{-1.41}) \times 10^{-3}$  & $38.99^{+0.09}_{-0.05}$ & 19.00/14 & 0.17 \\
         & XMM1 & $2.26^{+1.32}_{-0.97}$  & $< 2.86 \times 10^{-6}$& $38.48^{+0.22}_{-0.31}$&  2.73/3 & 0.44 & $ 0.49^{+1.09}_{-0.24}$ & $< 0.11$ & $38.30^{+0.19}_{-0.31}$ & 4.24/3 & 0.24   \\ 
         T2 & C2 & $2.13^{+0.38}_{-0.33}$ & $(4.40^{+0.88}_{-0.87}) \times 10^{-6}$ & $38.86^{+0.62}_{-0.91}$ & 7.20/4 & 0.12 & $0.66^{+0.21}_{-0.15}$ & $< 1.15 \times 10^{-2}$ & $38.67^{+0.08}_{-0.09}$ & 1.42/4 &0.84 \\

          & XMM1 & $1.72^{+0.42}_{-0.40}$& $(3.45^{+0.98}_{-0.98}) \times 10^{-6}$ & $38.87^{+0.14}_{-0.15}$ &  8.82/7 & 0.27 & 
          $1.11^{+0.88}_{-0.37}$ & $<2.22 \times 10^{-3}$ & $38.71^{+0.17}_{-0.16}$ & 13.54/7 & 0.06   \\ 
         & C3&$1.81^{+0.39}_{-0.33}$ & $(4.35^{+0.92}_{-0.90}) \times 10^{-6}$ & $38.94^{+0.08}_{-0.09}$ & 4.35/4 & 0.36 & $1.02^{+0.41}_{-0.30}$ & $< 2.93 \times 10^{-3}$ & $38.78^{+0.11}_{-0.12}$ & 4.69/4& 0.32\\
         & C4 & $2.64^{+1.47}_{-0.93}$ & $(1.69^{+0.91}_{-0.69}) \times 10^{-6}$& $38.41^{+0.40}_{-0.20}$ & 5.52/6 (C) & 44.2\% & $0.42^{+0.39}_{-0.16}$ & $< 4.41 \times 10^{-2}$& $38.15^{+0.18}_{-0.18}$ &4.98/6(C) & 38.2\%\\
         T3 & C1  & $1.94^{+0.38}_{-0.35}$ & ($1.92^{+0.42}_{-0.38}) \times 10^{-6}$ & $38.51^{+0.10}_{-0.11}$ & 26.65/15(C) & 74.8$\%$ &$0.50^{+0.22}_{-0.12}$ & $<1.73 \times 10^{-2}$& $38.31^{+1.57}_{-0.09}$ &26.21/15(C)  & 81.3$\%$ \\
         & C3 & $2.25^{+1.24}_{-1.01}$ & $(1.35^{+0.87}_{-0.63}) \times 10^{-6}$& $38.33^{+1.29}_{-0.17}$&  9.85/7(C) & 67.9$\%$ & $0.69^{+1.31}_{-0.31}$ & $ < 1.18 \times 10^{-2}$ & $38.09^{+0.25}_{-0.17}$& 11.94/7(C)&  82.6\%\\
         &C4 & $2.87^{+1.10}_{-1.80}$& $<9.19 \times 10^{-7}$ & $38.07^{+1.13}_{-0.59}$& 22.68/28(C) &97.2$\%$& $0.31^{+4.10}_{-0.14}$& $< 0.31$ & $37.80^{+0.26}_{-0.43}$&23.24/28(C) & 97.4$\%$ \\
         T4& C1 & $1.33 (f)$ & $< 3.45 \times 10^{-7}$ & $ < 38.04$& 7.38/8(C) & 38.3\% &$1.20 (f)$ & $(2.76^{+2.24}_{-2.49}) \times 10^{-5}$ &
         $37.57^{+0.26}_{-1.02}$&6.71/8(C) & 34.7$\%$ \\
         &C4 & $>1.07$ & $<4.93 \times 10^{-7}$ & $38.01^{+0.84}_{-1.18}$& 26.55/28(C) & 99.1\% &$<2.15$ & $< 21.29$ &         $37.81^{+0.86}_{-0.35}$&26.80/28(C) & 98.2$\%$ \\
         
         T5& C4 & $1.60^{+1.32}_{-1.13}$ & $(3.99^{+3.07}_{-2.74}) \times 10^{-7}$ & $37.93^{+0.40}_{-0.47}$ & 10.42/13(C) & 36.8$\%$   &$0.69^{+3.81}_{-0.39}$& $<5.84 \times 10^{-3}$ & $37.69^{+0.43}_{-0.40}$&9.37/13(C)& 36.6$\%$  \\
         T6 & C1& $2.59^{+0.34}_{-0.32}$ & $(2.80^{+ 0.53}_{-0.48}) \times 10^{-6}$ &$38.63^{+0.07}_{-0.08}$ & 51.30/11(C) & 99.3$\%$& $ 0.32^{+0.06}_{-0.04}$ & $(5.71^{+6.83}_{-3.31}) \times 10^{-2}$ &$38.49^{+0.07}_{-0.08}$ &40.68/12(C) &  98.6$\%$\\
         &C2 &$2.69^{+0.43}_{-0.42}$& $(3.25^{+0.69}_{-0.66}) \times 10^{-6}$ & $38.69^{+0.09}_{-0.10}$& 25.50/17(C) & 69.7\%&$0.29^{+0.08}_{-0.06}$ & $<0.27$ & $38.53^{+0.09}_{-0.10}$ &25.35/17(C) & 63.6\% \\
         T7 & C2 &$2.25^{+1.10}_{-0.82}$ & $(8.68^{+5.04}_{-4.10}) \times 10^{-7}$ & $38.14^{+0.19}_{-0.22}$ & 13.78/14 & 68.1\% &$0.77^{+0.76}_{-0.46}$  & $<3.54 \times 10^{-3}$ & $37.96^{+0.09}_{-0.25}$& 14.58/14(C)& 55.3\% \\
         &C3   &$2.44^{+1.10}_{-1.02}$ & $(6.99^{+3.50}_{-3.02}) \times 10^{-7}$ & $38.03^{+0.17}_{-0.23}$ & 4.15/5(C)& 37.5$\%$& $0.47^{+0.70}_{-0.22}$ & $< 2.98 \times 10^{-2}$ & $37.81^{+0.18}_{-0.23}$& 3.78/5(C)& 37.0$\%$ \\
         T8 &C1 &$2.77^{+0.31}_{-0.29}$ & $(2.97^{+0.54}_{-0.53}) \times 10^{-6}$ & $38.66^{+0.07}_{-0.08}$& 56.20/12(C)& 99.2\% &$0.27^{+0.05}_{-0.03}$ & $0.12^{+0.13}_{-0.07}$ & $38.53^{+0.06}_{-0.04}$ & 42.24/12(C) & 97.4$\%$ \\
         &C2 &$2.57^{+0.54}_{-0.50}$  & $(2.19^{+0.65}_{-0.60}) \times 10^{-6}$ & $38.52^{+0.12}_{-0.14}$ & 50.58/17 & 98.9\% &$0.25^{+0.07}_{-0.04}$ & $< 0.44$ & $38.42^{+0.13}_{-0.15}$ &49.89/17& 99.7\%  \\
         T9 & C4  &$2.08^{+1.18}_{-1.13}$  & $(3.39^{+2.52}_{-1.94}) \times 10^{-7}$ & $37.74^{+0.31}_{-0.32}$ & 7.84/16(C) & 8.2\% & $0.49 (f)$& $(1.04^{+0.74}_{-0.53}) \times 10^{-3}$ & $37.53^{+0.25}_{-0.29}$& 7.20/17(C) &6.9\%\\
        T10& C1&$2.31^{+0.21}_{-0.20}$& $(5.82^{+0.68}_{-0.65}) \times 10^{-6}$ & $38.96^{+0.05}_{-0.05}$&55.32/21(C)& 96.4\%& $0.36^{+0.06}_{-0.04}$& $(6.15^{+5.46}_{-2.99}) \times 10^{-2}$ & $38.77^{+0.04}_{-0.05}$& 38.66/12(C) & 90.9\%\\
         T11& C2& $3.13^{+4.90}_{-1.58}$& $<6.30 \times 10^{-7}$ & $37.72^{+1.33}_{-0.58}$ &11.25/10(C)&79$\%$& $0.29^{+0.49}_{-0.26}$ & $< 8.48 \times 10^{-2}$ & $37.56^{+0.36}_{-0.53}$&10.84/10(C)&65.0$\%$ \\
        T12&C2 & $0.85 (f)$ & $<2.83 \times 10^{-7}$ & $< 38.23$ & 6.28/13(C)& 2.4\% & $1.10 (f)$ & $< 7.04 \times 10^{-5}$ & $<37.82$ &6.62/13(C)&4.5\% \\    
         T13&C1& $1.17^{+0.75}_{-0.66}$& $(3.49^{+2.03}_{-1.58}) \times 10^{-7}$ & $38.13^{+1.09}_{-0.34}$&10.80/14(C)& 34.2\% & $1.53^{+0.19}_{-0.28}$& $<3.61 \times 10^{-4}$ & $37.91^{+0.34}_{-0.41}$&7.38/14(C)& 19.0\% \\
         &C2 & $1.14^{+1.07}_{-1.08}$& $(3.16^{+2.80}_{-2.28}) \times 10^{-7}$ & $38.14^{+0.37}_{-0.50}$& 16.06/13(C)& 84.1$\%$& $1.58^{+0.28}_{-0.35}$& $<4.16\times 10^{-4}$ & $37.90^{+0.52}_{-0.48}$&15.31/13(C)& 56.6\%\\ 
         &C4 & $0.63^{+0.99}_{-1.66}$ & $< 4.23 \times 10^{-7}$ & $38.11^{+0.39}_{-0.60}$& 9.50/10(C)& 66.5\%& $>1.08$& $< 5.73 \times 10^{-5}$& $38.06^{+0.32}_{-0.51}$   &9.46/10(C)& 76.5\% \\
         
         T14 & C1 &$0.66^{+0.93}_{-1.64}$ & $(3.57^{+2.91}_{-2.03}) \times 10^{-7}$ & $38.21^{+0.38}_{-0.37}$& 7.47/4(C) &78.9\% & $>1.11$& $<1.49\times 10^{-3}$ & $38.14^{+0.29}_{-0.44}$& 7.44/4(C)& 66.2\%  \\

         & XMM1 & $2.28^{+0.27}_{-0.26}$ & $(6.93^{+1.12}_{-1.12}) \times 10^{-6}$& $39.04^{+0.08}_{-0.09}$& 35.90/19 & 0.01 & 0.32 & $9.66 \times 10^{-7}$& $38.82$ & 63.77/19 &  - \\
 
         \hline
    \end{tabular}
    \label{Table 4}
    \begin{flushleft}
    \small
    Note - $^{a}$ Spectral models used for fitting: PL - Power law; DISKBB - Disk blackbody, $^{b}$ source name, $^{c}$ Observation IDs used, $^{d}$ power law index, $^{e}$ normalisation of power law, $^{f}$ logarithmic unabsorbed X-ray luminosity (in the 0.3-10 keV energy range) in units of $\rm erg~s^{-1}$; $^{g}$ $\chi^2$/d.o.f, (C) indicates that the C-statistics is used,$^{h}$Goodness of fit (Goodness when using C-statistics and null hypothesis probability for $\chi^2$ statistics), $^{i}$ Inner disc temperature in units of keV, $^{j}$ normalisation of DISKBB. In some cases, parameter estimations were not possible, thus we fixed the parameters, which are represented as "($f$)". In some sources, the lower/upper error estimation was not possible in XSPEC; thus, we quoted the upper/lower limits of the parameters. For the source T14, the reduced $\chi^2$ for the DISKBB model exceeds 2; hence, parameter error estimation is not possible for the XMM1 observation.  
    \end{flushleft}
\end{table*}

\begin{figure}
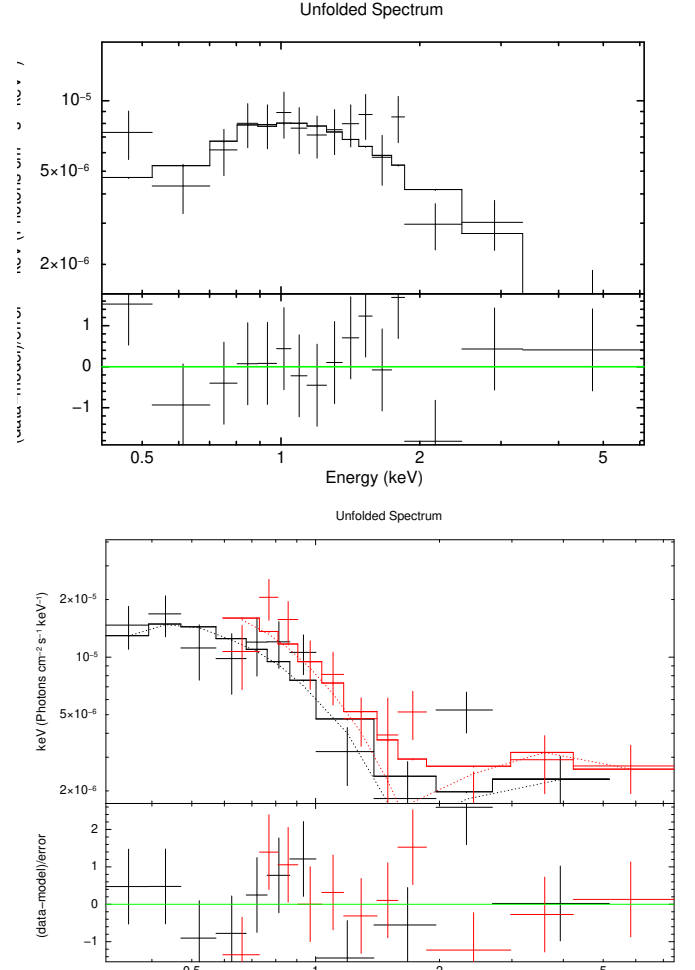

\begin{center}

\includegraphics[width=6.5cm,angle=-90]{T1_pl.eps}
\includegraphics[width=6.5cm,angle=-90]{t14_xmm1_bbody_diskbb.eps}

\caption{Spectral fit of transient and variable ULXs in NGC 4552. Top Panel: T1 spectrum from C1 observation is fitted with a power law model. Bottom Panel: T14 spectra from XMM1 observation fitted with blackbody plus disk blackbody model. In the XMM1 observation, the black and red data points represent the PN and combined MOS spectra, respectively.}

\label{spec_t1_t2_t14}
\end{center}
\end{figure}

Source T1 is one of the three brightest off-centre X-ray sources detected in this galaxy. This source is referred to as Src 41 in the previous work \citep{Xu_2005}. This source has 363 net counts in the C1 observation and was not detected in any other {\it Chandra} observations. A single component model, disk blackbody, provided an acceptable fit with inner disk temperature T$_{in}$ = $0.82^{+0.13}_{-0.11}$ keV, $\chi^2$/ d.o.f = 19/14. The estimated unabsorbed luminosity in the 0.3 - 10 keV band is $log~L_{X}$ =  $38.99^{+0.09}_{-0.05}\rm~erg~s^{-1}$. In the case of an absorbed power law model fit, the reduced $\chi^2 > 2$. Thus, we fitted the spectrum by considering the absorption column density as a free parameter. The best-fit value of $N_{H}$ is higher than the Galactic column density, and it is $2.87 \times 10^{21}$ cm\textsuperscript{-2} with $\Gamma$ = $2.67^{+0.49}_{-0.41}$, $\chi^2$ /d.o.f = 12.36/13. The unabsorbed luminosity (0.3-10 keV) of log $L_{X}$ = $39.40^{+0.20}_{-0.14} \rm~erg~s^{-1}$ (see Table \ref{Table 4}). We also fitted this source spectrum with a two-component model (DISKBB + BBODY), which provided a reasonable fit. The spectral parameters are given in Table \ref{Table 5}. The single power law provides the lowest $\chi^{2}$ among these model fits. The spectral fit of the T1 source from the C1 observation using the power law model is shown in the top panel of Figure \ref{spec_t1_t2_t14}. 

\begin{table*}
\caption {Best-fit parameters of the transient and variable ULXs using double thermal (DISKBB + BBODY) model}
    \centering 
    \setlength{\tabcolsep}{8.0pt}
    \begin{tabular}{ccccccccc}
    \hline

    $\rm Source^{a}$ & $\rm ObsID^{b}$ & $\rm {kT_{in}}^c$ & $\rm Norm_{Diskbb}^d$ & $\rm {kT_{BB}}^e$ & $\rm Norm_{BB}^f (10^{-7})$ & $\rm log L_{X}^{g}$ & ${{\chi^2}/d.o.f}^h$ & $\rm P_{null}^i$ \\ 
    \hline
    T1  & C1 & $2.93^{+2.70}_{-1.80}$ & $(1.62^{+116.50}_{-1.61}) \times 10^{-5}$ & $0.34^{+0.05}_{-0.03}$ & $2.09^{+0.56}_{-1.29}$&$39.07^{+0.14}_{-0.11}$ & 13.08/12 & 0.36  \\
    T14  & XMM1 & $0.25^{+0.05}_{-0.04}$ & $0.38^{+0.44}_{-0.21}$ & $1.39^{+0.98}_{-0.40}$ & $2.97^{+3.47}_{-1.12}$ & $39.13^{+0.13}_{-0.11}$ & 22.10/17 & 0.18  \\ 
 \hline
\end{tabular}
 \label{Table 5}
\begin{flushleft}
    \small
    Note - $^{a}$ Source name, $^{b}$ Observation IDs used, $^{c}$ Inner disc temperature in units of keV, $^{d}$ Normalisation of DISKBB, $^{e}$Black body temperature in keV, $^{f}$ BBODY Normalisation, $^{g}$ logarithmic unabsorbed X-ray luminosity in units of $\rm erg~s^{-1}$, $^{h}$ $\chi^2$/d.o.f, $^{i}$ the null hypothesis probability for $\chi^2$ statistics. 
    \end{flushleft}

\label{Table 5}
\end{table*}

Source T2 is another relatively bright source in the galaxy, detected in observations C2, XMM1, C3, and C4, but not in C1. From the power law fit, we found that for three out of four observations, T2 favours the absorbed power law model (based on $\chi^2$ values and null hypothesis probability) with a photon index $\Gamma$ $\sim$ 1.72 -- 2.64. The X-ray luminosity varies between $10^{38}$ - $10^{39} \rm erg~s^{-1}$. The diskbb model gives an inner disk temperature of 0.42-1.11 keV in these observations. The spectral modelling with a power law and its spectral parameter indicates that the T2 spectrum is marginally hard.

The source T14 was detected in the initial observations (C1 and XMM1) and disappeared in the later {\it Chandra} observations. This source has limited counts in C1 observations and is fitted with a power-law or disk blackbody model. In the XMM1 observation, the spectra cannot be explained by a single disk blackbody model ($\chi^{2}_{r} > 2$), but the power law model yields $\Gamma \sim 2.3$ with $\chi^2$/d.o.f = 35.9/19. The X-ray luminosity in the 0.3-10 keV band is $> 10^{39} \rm erg~s^{-1}$. We also fitted the {\it XMM-Newton} spectrum with a two-component model, which provided a better fit than the single power law model with a luminosity above $10^{39}\rm erg~s^{-1}$. The best-fit spectrum is shown in the bottom panel of Figure \ref{spec_t1_t2_t14}. The difference in the $\chi^{2}$ values between the single (power law) and double components models is 13.8 for the loss of two additional dof, which is significant at a confidence level of 98.4\% (from the F-test).

The remaining sources have limited counts available in the observations. Thus, it is difficult to conclude which model favours the spectrum more. However, when we modelled the spectra of these sources with power law or disk blackbody models, the luminosities of all sources were in the range of $10^{37} - 10^{38} \rm erg~s^{-1}$. Their spectral parameters are given in Table \ref{Table 4}. 

The sources given in Table \ref{Table 2} are not detected in some observations. For undetected cases, we derived upper limits on the count rate from all undetected observations using the APRATES command with a 90\% confidence level. It is then converted to flux in the energy range 0.3 - 10 keV using WEBPIMMS, using the best-fit model parameters. The flux is converted into luminosity by assuming a distance of 16.25 Mpc \citep{Tonry}. We have used the strict upper limit on the luminosity among the undetected cases to estimate the luminosity ratio. The strict upper limit on luminosity and ratio are given in Table \ref{Table 2}. This estimate is considered as the non-detection state luminosity and is used for the transient and variable source classification discussed in Section \ref{subsec:transient}.

\subsection {Potential NIR Counterparts}

\begin{figure*}
\centering
\includegraphics[width=\textwidth]{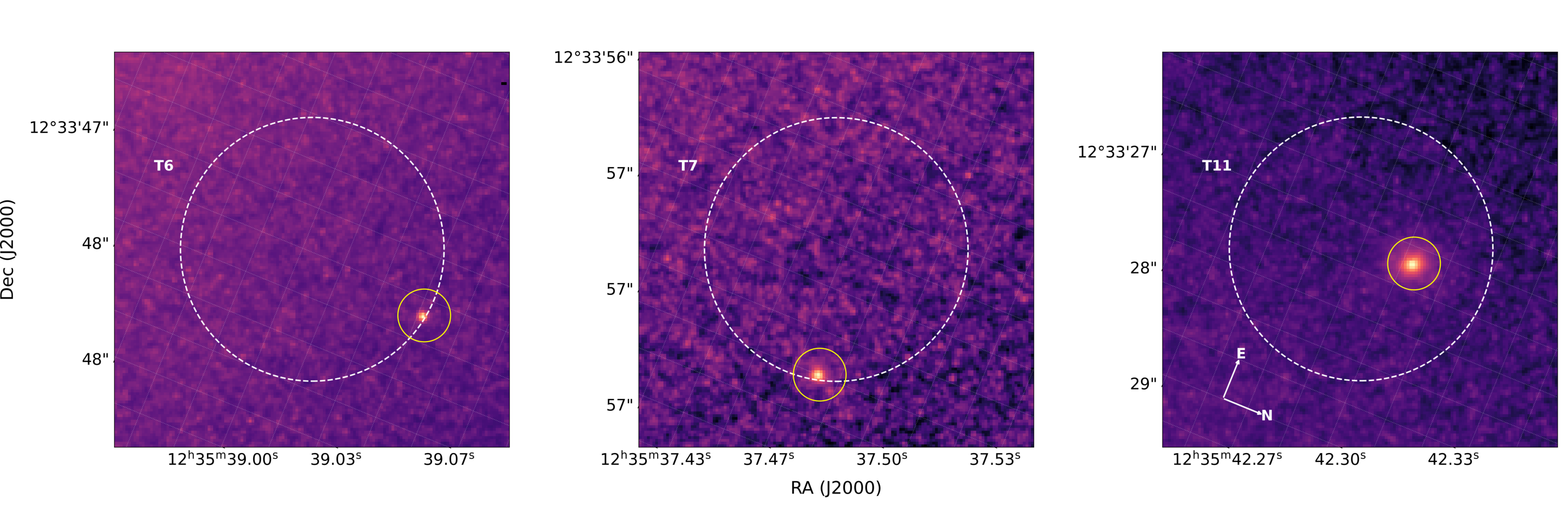}
\caption{$4.5 \times 4.5$ arcsec {\it NIRCam~F090W} image cutout around the X-ray positions of sources T6, T7, and T11 that have potential counterparts. The centre of the dashed white circle (1.5$^{\prime\prime}$ radius) represents the {\it Chandra} X-ray position, and the solid yellow circles (0.3$^{\prime\prime}$ radius) represent possible NIR counterparts detected at 3$\sigma$ threshold.}
\label{fig:Counterparts}
\end{figure*}

We perform point-source detection and aperture photometry on {\it JWST} images using the Python package \(photutils\)\footnote{\href{https://photutils.readthedocs.io/en/stable/index.html}{Photutils documentation}}. Source detection was performed with a detection threshold $3\sigma$. We then overlaid the X-ray positions of the transient and variable sources obtained from {\it Chandra} observation (listed in Table \ref{Table 2}) on the {\it JWST/NIRCam} mosaic image. We can see that only six out of 14 sources (T3, T6, T7, T8, T10 and T11) fall on NIRCam observations. However, in filters F090W and F150W, the source T3 falls on the chip gap. To search for potential NIR counterparts, we used a search radius of 1.5 arcseconds from the centre of the X-ray position. For three sources (T6, T7 and T11), we identified an NIR source in the 1.5 arcsecond search radius (see Figure \ref{fig:Counterparts}), and for the rest of the sources, no NIR sources were identified. However, these potential counterparts were not detected in some filters at the $3\sigma$ detection threshold, but were detected at the $2\sigma$ level. 

For these NIR sources, the aperture photometry was performed using different circular apertures with radii 1 -- 8 pixels (for F090W and F150W, 1 pixel = 0.031 arcsecond and for F277W and F356W, 1 pixel = 0.063 arcsecond). The background was extracted from a circular annulus with an inner radius of 9 pixels and an outer radius of 15 pixels. The Vega magnitude of the source is calculated from the flux density using the formula given on the \(JWST\) website
\footnote{\href{https://jwst-docs.stsci.edu/jwst-near-infrared-camera/nircam-performance/nircam-absolute-flux-calibration-and-zeropoints}{JWST Calibration and Zero points}}

\[
Vega_{\mathrm{mag}} = -2.5 \log_{10} \left( \frac{\mathrm{flux\ density} \times \mathrm{PIXAR\_SR}}{\mathrm{flux}_{\mathrm{Vega}}} \right)
\]

where the flux density is in $MJy/sr$ and the $flux_{vega}$ in $MJy$. The Vega flux is obtained from $jwst\_1126.pmap$\footnote{\url{https://jwst-docs.stsci.edu/files/216457093/216457096/1/1762453608965/NRC_ZPs_1126pmap.txt}} and the average area of a pixel in steradian $(PIXAR\_SR)$ from the FITS header. The estimated magnitude and colours are given in Table \ref{tab: Table 2}. The estimated magnitudes are not dereddened.

\begin{table*}
\centering
\caption{Vega magnitudes and colour indices of potential NIR counterparts}
\label{tab: Table 2}
\begin{tabular}{cccccccccc}
\hline
Source & 
F090W & 
F150W &
F277W &
F356W &
(F090W$-$F150W) & 
(F150W$-$F277W) & 
(F277W$-$F356W)
\\
\hline

T6 & $22.382 \pm 0.013$ & $20.956 \pm 0.017$ & $20.545 \pm 0.035$ & $20.098 \pm 0.042$ & $1.426 \pm 0.021$ & $0.411 \pm 0.039$ & $0.447 \pm 0.055$ \\
T7 & $22.210 \pm 0.007$ & $21.091 \pm 0.011$ & $20.621 \pm 0.025$ & $20.412 \pm 0.034$ & $1.119 \pm 0.013$ & $0.470 \pm 0.027$ & $0.209 \pm 0.042$ \\
T11 & $19.012 \pm 0.001$ & $17.792 \pm 0.002$ & $17.426 \pm 0.002$ & $17.247 \pm 0.003$ & $1.220 \pm 0.002$ & $0.366 \pm 0.003$ & $0.179 \pm 0.004$ \\
\hline
 
\end{tabular}%

\end{table*}

\section{DISCUSSION}
\label{sec:discussion}

We searched the transient and variable X-ray sources in the elliptical galaxy NGC 4552 using archival \textit{Chandra} and \textit{XMM-Newton} observations. Our analysis revealed that 14 sources in the optical extent of the galaxy showed peak-state and undetected flux behaviour, and thus identified them as transient and variable sources. Based on the peak-state and undetected luminosity ratio, we classified them as transient sources (2 sources), variable sources (1 source), and the remaining as mildly variable sources. The HR analysis of these sources suggests that most of them belong to the XRB category.

The source T1 is one of the bright sources in NGC 4552 and is about 90 arcsec from the centre. This source was designated "Src 41" in the previous study \citep{Xu_2005} and showed no evidence of variability during the observation. However, in this work, we identified this source as a transient. In the first {\it Chandra} observation (C1), the luminosity was above $10^{39} \rm erg~s^{-1}$, then dropped below $10^{39} \rm erg~s^{-1}$ in the {\it XMM-Newton} observation and was undetectable in the subsequent observations. From the upper-limits estimate of luminosity, it is clear that the source varied its intensity by more than an order of magnitude and became undetectable. The single component model fitting of T1 suggests a soft ($\Gamma \sim$ 2.7) or disk-dominated ($\rm kT_{in} \sim$ 0.8 keV) nature of the source. The nature of the transient sources in M31 has been examined using a double-thermal model \citep{2014ApJ...791...33B}, and they created a parameter space using $\rm kT_{in}$ and 2-10 keV luminosity to classify the transient sources as NS or BH X-ray binaries. Their analysis of 46 sources in M31 with disk blackbody and blackbody model parameters revealed that none of the sources occupied the NS LMXB parameter region, suggesting that they are black hole candidates. Following \cite{2014ApJ...791...33B}, we modelled the T1 spectrum using the double-thermal model (see Table \ref{Table 5}), but the spectral fit did not improve significantly over the single power law model. However, the yielded inner disk temperature ($\rm kT_{in} \sim$ 2.9 keV) and the 2--10 keV luminosity $\sim 6.2 \times 10^{38} \rm erg~s^{-1}$ suggest NS LMXB nature \citep{2013ApJ...772..126B} for this source, although the reported luminosity is too high for typical NS LMXBs. However, these luminosity levels can be exhibited by accreting BHs. The earlier {\it Chandra} X-ray spectral analysis inferred the source as a black hole candidate with a mass of 15-135 $M_\odot$, depending on the spin of the system \citep{Xu_2005}. Moreover, the source T1 is in the joint {\it Chandra}-{\it Hubble Space Telescope} ({\it HST}) field, and optical analysis with {\it HST} suggests that the source is associated with a globular cluster \citep{Xu_2005}. The bright X-ray source and globular cluster association has been seen in other early-type galaxies, and many of the X-ray sources belong to the ULX category \citep{2001ApJ...557L..35A, 2003ApJ...585..756J, 2019MNRAS.485.1694D, 2023MNRAS.518.3386T}. 


 
The source T14 is relatively faint in C1 observations with the luminosity of $\sim 1.6 \times 10^{38} \rm erg~s^{-1}$ and became bright in the 2003 {\it XMM-Newton} observation, where the luminosity changed by at least a factor of $\sim 7$ and emitting just above $10^{39} \rm erg~s^{-1}$, when fitted with single power law model. The dual thermal model fitting also estimated a luminosity above $10^{39} \rm erg~s^{-1}$. The spectral results ($\rm kT_{in} \sim$ 0.25 keV and the 2--10 keV luminosity $\sim 6.8 \times 10^{38} \rm erg~s^{-1}$) using the model combination prescribed in \cite{2014ApJ...791...33B} suggest that the source may be a black hole candidate. However, given the limited number of counts for the source, it is difficult to pinpoint the nature of the compact object.

Many ULXs detected in nearby galaxies are persistent in nature \citep{2009MNRAS.397.1836G, kaaret2017ultraluminous}. However, a new class of transient ULXs has been identified in nearby galaxies \citep{2010ApJ...710L.137F, 2012ApJ...750..152S, 2019MNRAS.483.3566V, 2020ApJ...891..153E, 2022MNRAS.511.4528U, 2022MNRAS.515.4669R, 2023ApJ...951...51B, Allak2023, 2024MNRAS.528..418L}. These sources exceed $10^{39} \rm erg~s^{-1}$ in luminosity and exhibit transient behaviour. The two sources in NGC 4552, T1 and T14, show these features in the archival observations, and we classify them as transient and variable ULXs. The transient ULXs exhibited both soft \citep{2022MNRAS.515.4669R, Allak2023} and hard spectral characteristics  \citep{2019MNRAS.483.3566V, 2020ApJ...891..153E}. However, the transient and variable ULXs in NGC 4552 (T1 and T14)  exhibited a soft spectral nature. ULXs exhibited spectral turnover at higher energies \citep{2006MNRAS.368..397S, 2009MNRAS.397.1836G}, which is an indication of super-Eddington accretion. Due to the limited counts in the spectra, we were unable to ﬁnd evidence of the spectral turnover at high energies. However, the significant variability in terms of luminosity and the soft spectral nature appear to be consistent with some of the super-Eddington ULXs \citep[e.g.,][]{2012MNRAS.420.1107P, 2017ApJ...839...46S, 2017ApJ...839..105W}. It is obvious from the analysis that we have only one observation with a luminosity above the threshold for ULXs, and the nature of the accretor in both sources is not clear from the current data. Transient and variable ULXs are transitional objects between normal XRBs and ULXs. For example, in the nearby galaxy NGC 55 the previously detected transient X-ray source \citep[XMMU J001446.81-391123.48;][]{jithesh2015x} with a luminosity around a few $10^{38} \rm erg~s^{-1}$ in the 2010 XMM-Newton observation exceeds the luminosity threshold $10^{39} \rm erg~s^{-1}$ in the new, deeper XMM-Newton observation, and identified the source as a new transient ULX in that galaxy \citep{2022MNRAS.515.4669R}. Thus, searching for transient and variable X-ray sources is essential to understanding the new population of ULXs in external galaxies. 


The source T2 appeared as a new source after the first {\it Chandra} observation (C1) and was detected in all subsequent observations. The source luminosity has varied by at least a factor of $\sim 15$ between the detected and undetected observations. The hardness ratio analysis indicates that the source belongs to the XRB category. The spectral modelling with a power law component showed a marginal variation in the power law index, with large uncertainties, and the luminosity varied by a factor of 3 across the detected observations. For the remaining sources, the spectral analysis indicates that they may be in the hard state or disk-dominated state \citep{2006ARA&A..44...49R}. For the majority of these sources, the X-ray luminosity is $< 10\%$ of the Eddington luminosity for NS or BH X-ray binaries. The limited counts in the detected spectra pose a challenge for the detailed modelling of these sources. However, these sources may belong to the neutron star or black hole X-ray binary category solely on the basis of luminosity, and longer observations are required for further classification.

Supersoft sources \citep{1997ARA&A..35...69K} generally contribute to $\sim 25-35 \%$ of the transient source population in nearby galaxies \citep{2008ApJ...680.1120W, 2009A&A...500..769H, Jithesh_2016, Jithesh_2017}. However, we did not find any supersoft sources in NGC 4552, which may be due to the large distance of the galaxy (16.25 Mpc) and the limited sensitivity of the observations \citep{2010NewAR..54...72D, 2021A&A...646A..85G}. 

We searched for potential counterparts of the transient and variable sources in the {\it JWST} observations. The JWST/NIRCAM filters utilised in this study have angular resolution of 0.035” (F090W), 0.058” (F150W), 0.106” (F277W) and 0.138” (F356W), which corresponds to spatial resolution of 2.76 pc, 4.57 pc, 8.35 pc, and 10.87 pc, respectively, at the distance of NGC 4552. Given these spatial scales, JWST cannot resolve individual stars from compact stellar associations or small star clusters in NGC 4552. Therefore, the identified near-infrared counterparts should be interpreted either as a single luminous star or a compact star cluster. Six sources fall in the field of view of {\it JWST} images, among them, three X-ray sources have potential NIR counterparts. Unfortunately, the transient and variable ULXs are not in the {\it JWST} field of view. The yielded absolute magnitudes of two mildly variable sources (T6 and T7) range from $-8.7$ to $-10.9$, and for T11, the obtained magnitudes are $-12.0$ to $-13.8$. The NIR counterparts of LMXRBs, neutron star X-ray binaries and ULXs have been identified in the past  \citep{Mauerhan2009, Homan2009, Curran2011, Heida2014, Shaw2017, Lopez2017, Allak2023}. The NIR counterparts of seven ULXs in different galaxies within 10 Mpc have been identified \citep{Lopez2017}. The reported absolute magnitudes range from -9.3 to -11.2, consistent with those of red supergiants (RSGs). In another study of NIR counterparts of ULXs, \cite{Heida2014} identified NIR sources with absolute magnitudes ranging from -8.1 to -11.3 in the K-band, consistent with RSGs. For RSGs, the absolute magnitude in the K band ranges from $-8.0$ to $-10.5$ \citep{Elias1985, Drilling2000}. For the two potential NIR counterparts (T6 and T7), the obtained absolute magnitudes are consistent with the aforementioned ranges \citep{Elias1985, Drilling2000, Heida2014, Lopez2017}, suggesting that the counterparts may be RSGs. However, for T11, the absolute magnitude ($-12$ to $-13.8$) is too bright for a single RSG \citep{Heida2014}. The calculated X-ray to NIR flux ratio $\sim 10^{-3}$ (for T6 and T7) and $\sim 10^{-5}$ (for T11), which are consistent with Galactic X-ray sources, ecliptic X-ray binaries, and other X-ray sources \citep{Ananth1984}.

The transient and variable sources identified in this work were cross-matched with X-ray and multiwavelength source catalogues. The sources reported in the work were listed in the Chandra source catalogue Release 2 (CSC 2.1; \citealp{Evans}). Among these, T1 is listed in the Expanded ULX catalogue \citep{Bernadish}. However, none of the other sources was identified in any major ULX catalogues, such as the 2XMM ultraluminous X-ray source candidates  \citep{Walton11}, the 4XMM-DR10/CSC2/2SXPS ULX candidates \citep{Walton22}, and the ultraluminous X-ray candidates \citep{swartz4}. In addition, we searched for the possible counterparts of these sources in multiwavelength catalogues, such as 2MASS \citep{2mass}, ALLWISE \citep{Allwise}, and catWISE 2020 \citep{catwise}, optical source catalogues such as Gaia DR3 (epoch 2016; \citealp{Gaia}), SDSS DR16 \citep{sdss}, and PanSTARRS Release 1 DR2 \citep{panstarrs}, and AGN catalogues such as ALLWISE AGN Catalogue \citep{AGN_ALLWISE}, Milliquas Catalogue \citep{2023OJAp....6E..49F} and Veron-Cetty \& Veron Quasar Catalogue \citep{Veron}. No counterparts were found in the AGN catalogues of \citet{AGN_ALLWISE}, \citet{2023OJAp....6E..49F}, and \citet{Veron} within the search radius of 1.5 arcseconds. Within a $1.5^{\prime\prime}$ matching radius, no counterparts were listed in IR source catalogues such as 2MASS \citep{2mass}, and catWISE 2020 \citep{catwise}. Possible optical counterparts were identified for T1 in PanSTARRS Release 1 DR2 \citep{panstarrs} and for T11 in Gaia DR3 (epoch 2016; \citealp{Gaia}), PanSTARRS Release 1 DR2 \citep{panstarrs}, and SDSS DR16 \citep{sdss} within $1.5^{\prime\prime}$. The possible counterpart of T1 identified in the PanSTARRS catalogue has the AB magnitudes of 21.41, 20.88, 20.67, 20.63 and 19.69 in the g, r, i, z, and y bands, respectively. For T11, the PanSTARRS source has an angular separation of $0.56^{\prime\prime}$ from the X-ray position, with AB magnitudes of 19.64, 18.96, 18.68, 18.66 and 18.30 in the g, r, i, z and y bands, respectively. The above-mentioned T11 PanSTARRS source is also listed in the Gaia DR3 catalogue, with a g-band magnitude of $20.43$, and the SDSS DR16 catalogue with AB magnitudes of $12.68$ (r-band) and $14.48$ (g-band). We note that the position of the T11 counterpart identified in the PanSTARRS, Gaia DR3 and SDSS DR16 catalogues is consistent with the JWST counterpart identified in this work.  

To summarise, our study identified the transient and variable X-ray source population in NGC 4552 using the archival {\it Chandra} and {\it XMM-Newton} observations. The population comprises 14 sources, most of which are consistent with XRBs. Among them, two sources exceed the luminosity $10^{39} \rm erg~s^{-1}$ in detected observations, and we classified them as new transient and variable ULXs in NGC 4552. The transient ULX is identified in the joint {\it Chandra}-{\it HST} field, and the counterpart coincides with a globular cluster. The NIR counterparts of three mildly variable sources are consistent with RSGs. This study added a couple of sources to the growing population of transient ULXs, and deep monitoring observations of these sources can shed further light on the nature of the new transient and variable ULXs in NGC 4552.

\section*{Acknowledgments}

We thank the anonymous referee for constructive comments that have improved the manuscript. This work has made use of data obtained from the High Energy Astrophysics Science Archive Research Center (HEASARC), provided by NASA’s Goddard Space Flight Center. This research has used data obtained from the Chandra Data Archive and the Chandra Source Catalog, and software provided by the Chandra X-ray Center (CXC) in the application package CIAO. This research is based on observations obtained with XMM-Newton, an ESA science mission with instruments and contributions directly funded by ESA member states and NASA. This work is based [in part] on observations made with the NASA/ESA/CSA James Webb Space Telescope. The data were obtained from the Mikulski Archive for Space Telescopes at the Space Telescope Science Institute, which is operated by the Association of Universities for Research in Astronomy, Inc., under NASA contract NAS 5-03127 for JWST. We acknowledge the use of Grammarly for proofreading the paper. VJ acknowledges the Centre for Research Projects in Sciences, CHRIST (Deemed to be University), for the financial support provided through a Seed Money Grant (CU-ORS-SM-24/24). VJ thanks the Inter-University Centre for Astronomy and Astrophysics (IUCAA), Pune, India, for the Visiting Associateship.

\section*{Data Availability}

The data used in this article are available in the HEASARC database 
\href{https://heasarc.gsfc.nasa.gov}{(https://heasarc.gsfc.nasa.gov)}. The results reported in this article are based on archival observations made by the James Webb Space Telescope, obtained from the data archive at the Space Telescope Science Institute \href{https://mast.stsci.edu/portal/Mashup/Clients/Mast/Portal.html}{(https://mast.stsci.edu/portal/Mashup/Clients/Mast/Portal.html)}



\bibliographystyle{mnras}
\bibliography{bibtex} 




\appendix




\bsp	
\label{lastpage}
\end{document}